\documentclass[preprint,5p,twocolumn,authoryear]{elsarticle}

\usepackage{amssymb}
\usepackage{amsmath}
\usepackage[colorlinks = true,
  linkcolor = blue,
  citecolor = blue,
  urlcolor = blue]{hyperref}
\usepackage{balance}
\usepackage{graphicx}
\usepackage{booktabs}
\usepackage{multirow}
\usepackage{tabularx}
\usepackage{enumitem}
\usepackage{threeparttable}
\usepackage{url}

\usepackage{fancyhdr} 
\fancypagestyle{firststyle}
{
   \fancyhf{}
   \fancyfoot[C]{\scriptsize{\textit{Computers \& Security}, 2025, vol.~159, pp.~104689. This version is the authors' copy. \\ The publisher's definite version is available at Elsevier via \url{https://doi.org/10.1016/j.cose.2025.104689}.}}
}

\begin{document}

\title{From Cyber Security Incident Management to \\ Cyber Security Crisis Management in the European Union}
\author[sdu]{Jukka Ruohonen\corref{cor}}
\ead{juk@mmmi.sdu.dk}
\author[utu]{Kalle Rindell}
\author[uot]{Simone Busetti}
\cortext[cor]{Corresponding author.}
\address[sdu]{University of Southern Denmark, Denmark}
\address[utu]{University of Turku, Finland}
\address[uot]{University of Teramo, Italy}

\begin{abstract}
Incident management is a classical topic in cyber security. Recently, the
European Union (EU) has started to consider also the relation between cyber
security incidents and cyber security crises. These considerations and
preparations, including those specified in the EU's new cyber security laws,
constitute the paper's topic. According to an analysis of the laws and
associated policy documents, (i)~cyber security crises are equated in the EU to
large-scale cyber security incidents that either exceed a handling capacity of a
single member state or affect at least two member states. For this and other
purposes, (ii)~the new laws substantially increase mandatory reporting about
cyber security incidents, including but not limited to the large-scale
incidents. Despite the laws and new governance bodies established by them,
however, (iii)~the working of actual cyber security crisis management remains
unclear particularly at the EU-level. With these policy research results, the
paper advances the domain of cyber security incident management research by
elaborating how European law perceives cyber security crises and their relation
to cyber security incidents, paving the way for many relevant further research
topics with practical relevance, whether theoretical, conceptual, or empirical.
\end{abstract}


\maketitle
\pagestyle{empty} 

\section{Introduction}

\thispagestyle{firststyle} 

``Europe has been in crisis management mode for a decade and a half''
\citep[p.~1]{Handler24}. Although the statement quoted is true in many respects,
it does not (yet) apply to the cyber security domain. In other words, to the
best of the authors' knowledge, there has not been a truly large-scale,
pan-European cyber security crisis thus far. This statement does not mean that
the EU would not have prepared for such a crisis. These preparations, including
those mandated by the EU's new cyber security laws, are what the paper is
about. The paper's relevance and contributions can be elaborated with four brief
points.

To start from practical relevance, first, the topic is timely. In fact, the
European Commission (EC) just recently released a new proposal for handling
large-scale, cross-country cyber security crises in Europe~\citep{EC25b}. The
second point follows from this timeliness; again to the best of the authors'
knowledge, the paper is the first to elaborate cyber security crisis management
in the EU context. While there are some existing works, they have concentrated
on national cyber security crisis management~\citep{Boeke17, Collier16,
  Ostby20}. That is, the EU context has been missing thus far, including with
respect to the EU's new judicial framework. In addition, governance below
national administrations has received limited attention in
general~\citep{Belad24}. Subsequently, third, the paper contributes to the
recent efforts to digest and analyze the EU's many new cyber security laws and
their implications~\citep[among many others]{Alexopoulos25, Busetti25, Mueck25,
  Rataj25, Ruohonen24I3E, Ruohonen25RE}. Regarding existing research and
knowledge more generally, fourth and last, the paper contributes to the incident
management research domain by elaborating, analyzing, and theorizing on how
incidents are related to and may transform into crises. Because crises typically
have immense consequences, the paper's overall relevance is well-justified; it
is important to understand how the EU's new cyber security legal framework is
designed with crisis management in mind.

With respect to the last point, both incident management research and frameworks
for it have traditionally operated at the organizational level; incidents are
something that organizations may face, and thus they should also prepare for
them. For instance, a classical incident management framework from the National
Institute of Standards and Technology (NIST) defines cyber security incident as
``a violation or imminent threat of violation'' of any cyber ``security
policies, acceptable use policies, or standard security practices'' established
and enforced by organizations~\cite[p.~6]{NIST12}. The organizational focus is
thus clear. According to an early literature review, a similar focus is present
in many other well-known frameworks and standards for incident management and
closely related topics~\citep{Tondel14}. From the previous quotation it also
follows that an organization should have cyber security policies established and
enforced because otherwise an incident remains only implicit and vaguely
defined. A similar claim applies to cyber security crisis management; also
larger entities, whether industry sectors, geographic regions, or countries and
beyond, should have policies in place for handling cyber security crises. As
will be seen, in the EU these policies are partially but still explicitly
written in recent cyber security laws.

A further point about the organizational focus is that it has typically been
present also in different socio-technical frameworks for incident management and
cyber security management in general~\citep{AlSabbagh15, Jaatun09,
  vanHaastrecht21}. Motivated by a recent work on layered cyber security
frameworks~\citep{Panteli25}, the paper adopts a multi-level perspective
sometimes used in the socio-technical frameworks~\citep{Malatji19}, but extends
it beyond organizations. Thus, the first part in the socio-technical term refers
to an adjective societal and particularly a noun European. Although societal
cyber security is not a well-defined concept, it has sometimes been discussed in
a context of critical infrastructure protection~\citep{Gjesvik19}. Such a
framing serves also the paper's purposes well because among the EU laws
considered is an important new law for critical infrastructure protection in
Europe.

With these motivating remarks in mind, the following three research questions
(RQs) are examined:

\begin{itemize}
\item{RQ.1: What constitutes a cyber security crisis in the European
    Union?}
\item{RQ.2: What do recent EU laws impose upon cyber security incident
  and crisis management?}
\item{RQ.3: How cyber security crises are managed, governed, and coordinated in
  the EU and by whom?}
\end{itemize}

The paper's remainder is structured into four sections. The opening
Section~\ref{sec: background} motivates the background further, including with
respect to the multi-level approach pursued and the EU laws considered. The
methodology for analyzing these is also elaborated, and a few clarifying
framings are also done to restrict and limit the paper's
scope. Section~\ref{sec: incident types} continues by elaborating how different
law-imposed incident types are related to the multi-level approach. In general,
the section answers to RQ.1 and RQ.2. The subsequent Section~\ref{sec: crisis
  management} answers to RQ.3 by elaborating the various institutions,
organizations, and networks involved in cyber security crisis management in the
EU context. The last three Sections~\ref{sec: conclusion}, \ref{sec:
  limitations}, and \ref{sec: further work} present a conclusion as well as a
few points about limitations and associated future research directions.

\section{Background}\label{sec: background}

\subsection{Framings}\label{subsec: framings}

Some framings are required for limiting and aligning the paper's scope to a
manageable composition. To begin with, the paper is framed toward cyber security
alone. The paper's opening quotation about Europe having been in a crisis
management mode for a long time helps to understand that crises vary a lot; from
financial crises all the way to wars and the existential climate change
crisis. While notions such as cascading effects and risks~\citep{Adkins20,
  Ruohonen24ISJGP, Ruohonen25RE} make it understandable that many crises may be
interconnected to each other, no attempts are made to connect cyber security
crises to other crises. The same applies regarding threats. Some cyber security
threats, including particularly those related to advanced persistent threat
(APT)~actors, are today, in Europe, often perceived to be linked with a broader
class of so-called hybrid threats~\citep{Anagnostakis23, Jungwirth23}. While
acknowledging the linkage's theoretical and practical validity, no attempts are
made to move beyond plain cyber security. By implication, as hybrid threats may
involve also a military dimension, also the EU's other cyber security
pillars~\citep{Christou16, Ruohonen24I3E}, whether cyber defense or data
protection, are excluded from the paper's scope. Furthermore, theorizing about
different types of cyber security crises are further omitted for brevity,
clarity, and alignment with the EU's laws considered.

With respect to large-scale cyber security crises, \citet[p.~12]{ENISA24a}
emphasizes a need to distinguish between a creeping crisis, ``which simmers
under the radar'' before ``suddenly and unexpectedly erupting'', an acute
crisis, ``which is sudden, unforeseen and can have a massive impact in a very
short amount of time'', and a recurring crisis, which occurs almost
continuously. Similar categorizations have been presented in the academic
literature~\citep{Boin18, Head22}. Of these three types, denial of service
attacks, in particular, could be seen as something recurrent, whereas a severe
data breach or a successful ransomware attack might be seen to belong to the
category of acute crises. Though, these examples are more about incidents than
crises; therefore, also risk analysis is presumably more challenging when trying
to assess crises rather than incidents; when operating at a level of societies
or beyond rather than at the conventional organizational level. At the societal
level, an example of a creeping crisis might involve a discovered large-scale
espionage campaign conducted by an APT actor. Beyond these brief remarks, as
said, more elaborate theorization is left for further work.

\subsection{Methodology}\label{subsec: methodology}

The paper operates in the domain of policy studies. Within this domain, a
separation between policy analysis and policy (process) research is sometimes
done. Broadly speaking, the former is about prescriptive research seeking to
inform policy-making, whether in terms of evaluations, impact assessments, or
something else, while the latter is descriptive research focused particularly on
theory-building~\citep{DeLeon10, Secchi16}. Given this characterization, the
paper is about policy research. Regarding descriptive policy research,
an~\textit{ex~post}, a~\textit{post hoc}, and generally a retrospective approach
is used~\citep[cf.][]{Patton16}. This choice is also unavoidable because the EU
laws considered have already been enacted. As soon described in
Subsection~\ref{subsec: strategy}, these laws also provide the primary materials
for the study.

Regarding \textit{ex ante} policy analysis, it is worth remarking that the EC
did prior impact assessments for most---but not all~\citep{EP24a}---of the laws
considered in the paper. For the paper's purposes, it is particularly worth
remarking that these assessments indicated a lack of joint situational awareness
and crisis management between the member states, and between them and the
EU-level administration; information sharing was concluded to operate on an
\textit{ad hoc} basis and cross-border spillovers were omitted from risk
analyses~\citep[pp.~21--22]{EC20a}. This point justifies an academic expert
evaluation of the three RQs postulated.

Regarding theory-building, the paper analyzes and theorizes how the EU's new
laws align with common analytical frameworks used in incident and crisis
management research. On the side of engineering and computer science, within
which theory-building often has a different meaning, the paper aligns with
requirements engineering research within which legal requirements are a distinct
genre~\citep{Ruohonen25RE}. This research branch justifies RQ.2. In other words,
it is important to know what laws require, whether from organizations,
producers, or public administrations. Once an answer is known, it is possible to
continue toward empirical policy research; without knowing the legal
requirements, formulation of relevant research questions is difficult---if not
questionable altogether. This point also applies to the framings done; insofar
as policy analysis is concerned, it is pointless to evaluate responses to hybrid
threats without knowing about the preparations and frameworks for handling them.

In addition, a brief scenario analysis is presented to better elaborate the
legal interpretations drawn and their potential practical implications. By
following existing research~\citep{Ruohonen25IST}, the short scenario analysis
presented is purely hypothetical, using an imaginary but still plausible case
for the elaborations. Although most scenario analyses are about foresight and
the future \citep{Sus20}, the analysis pursued aligns with the descriptive tenet
in policy studies. Besides elaborating the legal aspects, the scenario analysis
also helps at the theory-building tenet, including with respect to the
analytical levels subsequently described.

\subsection{Analytical Levels}\label{subsec: levels}

The paper's conceptual and theoretical background can be motivated by
considering an analytical relation between incidents and crisis through
different vertical levels of analysis often used in social sciences. A term
multi-level analysis is sometimes used to refer to approaches that operate at
two or more such levels~\citep{Belad24, Hutzschenreuter20}. A term multi-level
governance is a common alternative \citep{Bartle16}. The terms should not be
equated to multi-level (cyber) security, although there is a rough conceptual
similarity because both involve a hierarchy~\citep[cf.][]{Anderson20}. In any
case, an incident that involves multiple analytical levels could be called a
multi-level \text{incident---or}, depending on which levels an incident has
transcended, a cyber security crisis.

\subsubsection{The International Level}

The international level is at the top of the hierarchy. When operating at this
level, an analysis typically functions horizontally, focusing on relations
between states. In terms of cyber security, good examples about such interstate,
transnational relations would cover cyber norms, cyber diplomacy, and even what
is known and debated as cyber war. However, as said, such topics are beyond the
paper's scope. Another point is that it is also possible to consider information
systems and other technical solutions at this level; such systems and solutions
are those that transcend both national and organizational
boundaries~\citep{Rukanova15}. Such transcendence can also been seen as
something that separates cyber security incidents from cyber security crises.

\begin{figure}[th!]
\centering
\includegraphics[width=\linewidth       , height=4.5cm]{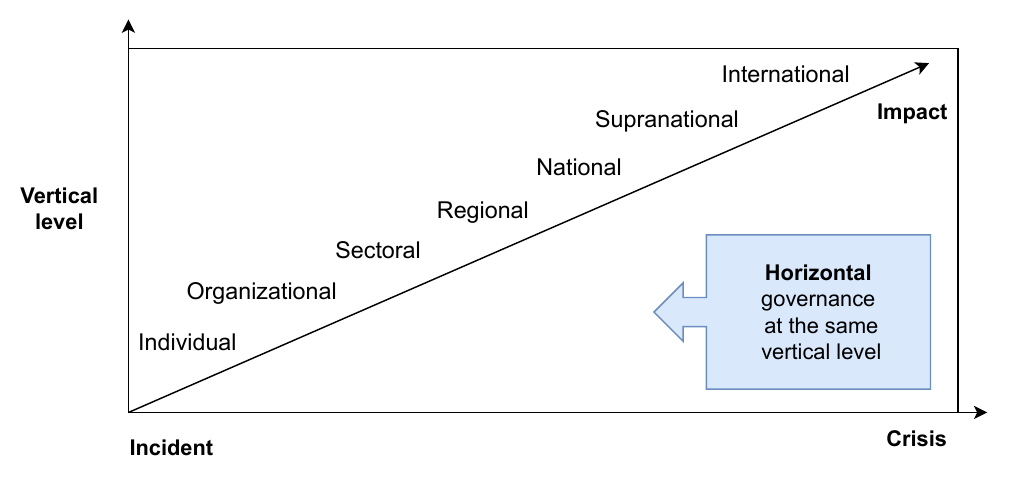}
\caption{Analytical Levels (adopted from \citealt[pp.~17--18]{ENISA24a})}
\label{fig: levels}
 \end{figure}

If a cyber security incident escalates to the international level, it is not
really any more a mere incident but rather a crisis. With this point in mind,
Fig.~\ref{fig: levels} displays seven analytical levels through which incidents
may become crises. As can be seen, the impact is taken to increase the further a
cyber security event is from an incident.

\subsubsection{The Supranational Level}\label{subsec: supranational level}

Below the international level is a supranational level. This level is at which
the EU operates; the European Union is a supranational union. At the same time,
however, it operates in intergovernmental terms; the Council of the European
Union would be the prime example in this regard. In contrast, the European Union
Agency for Cybersecurity (ENISA) operates at the supranational level, although
intergovernmental tenets may still be present in terms of staffing and related
aspects. While there is a classical debate regarding the meaning of and
interplay between these two theoretical governance terms~\citep{Bickerton22,
  Handler24, Tsebelis01}, without a particular loss of specificity and rigor,
the term supranationalism is used in case there is no particular reason to
emphasize the intergovernmental dimension.

Note also a difference between the terms transnational and intergovernmental;
the latter is about relations between governments, as is the case in the Council
in which ministers or other heads of states represent their member states,
whereas the former can be seen to cover also supranational governance because
the meaning is essentially about relations between nations. These relations may
be about states, but they may also include transnational non-governmental
organizations, different transnational collaboration and coordination networks,
and international organizations~\citep{Ganzle22}. When such transnational
relations are accounted for, the term multi-level governance can be seen to
cover also a horizontal dimension; a coordination and collaboration between
different entities not directly linked to governmental representatives. These
terminological clarifications are important because the EU's overall multi-level
crisis management framework covers numerous actors operating at different
governance levels and policy areas~\citep{Anagnostakis23}. The concepts also
provide a theoretical foundation for analyzing the EU's cyber security crisis
management framework in terms of governance and coordination.

\subsubsection{The National, Regional, and Sectoral Levels}\label{subsec: national regional and sectoral levels}

When taking a step downwards from the supranational level, the national level is
encountered. Regarding governance, authoritative computer security incident
response teams (CSIRTs) are particularly relevant in Europe; according to both
old and new EU laws, each member state should have an authoritative CSIRT or a
related competent authority for nationwide incident management.\footnote{~Here,
the adjective authoritative is used for emphasizing a difference to other
CSIRTs, including sectoral teams as well as teams within organizations and
companies. Note also that in practice the abbreviation CSIRT is equivalent to an
abbreviation CERT, denoting a computer emergency response team.} However,
presently it remains still somewhat unclear how national administrations are, or
will be, arranged in each and every member state. By hypothesis, the divergence
is, or will be, large across the twenty-seven member states.

Although there are no clear demarcations, operations at the national level are
sometimes (but not always) about crisis management rather than incident
management. The same could be said even about the regional level, especially
when keeping in mind that it includes both smaller geographically bound
administrative units as well as larger geographic units, such as is the case in
the Germany's federalist administration. Also crisis management tends to be
decentralized to the regional, \textit{L\"ander}, level in Germany, although
strategic management and high-level political coordination occurs at the federal
level~\citep{Christensen16}. Without the federalist twist, the setup is rather
similar to many other European countries within which strategic and political
aspects are oftentimes centralized, while the operational level is often more or
less decentralized~\citep{Christensen16, Gjesvik19, Ramsell12}. Regions also
intervene with industry sectors in many member states. Whether an example is
about the Ruhr region in Germany or the Emilia-Romagna region in Italy,
specialized industry activity has often concentrated to specific regions in
Europe and also elsewhere. Authoritative CSIRTs in some European countries have
also been decentralized with sectoral specialization in mind~\citep{Boin18}. In
terms of incident management, many industries also have different sectoral and
regional coordination networks and governance hubs.

\subsubsection{The Organizational and Individual Levels}

Then, below the sectoral level is the organizational level. As said in the
introduction, this level has been the traditional focal point in incident
management and its research. Again, coordination may occur horizontally between
organizations (inter-organizational coordination) and within organizations
(intra-organizational coordination), and vertically between organizations and
individuals~\citep{Heine19, Panteli25}. The intra-organizational coordination is
relevant also for establishing organizational CSIRTs; typical choices for large
organizations include outsourcing, a centralized organization-wide team, or
distributed teams across an organization's departments~\citep{Mitropoulos06,
  NIST12}. Organizations may also coordinate vertically with sectoral or
regional entities, such as industry associations, as well as entities operating
at a national level, including authoritative CSIRTs in particular. Finally, as
seen from Fig.~\ref{fig: levels}, it is also possible to consider incidents at a
level of individuals, whether employees, citizens, or consumers. Given the
extensive research and evidence on the human factors in cyber security, it is
also possible that an incident affecting an individual escalates to the
organizational level or even beyond that.

\subsubsection{An Example of a Multi-Level Incident}\label{subsec: danish}

A good example about a multi-level incident would be the attacks against
Danish energy sector companies in 2023. Given that over twenty energy sector
companies were compromised, the incident prompted also international media
attention~\citep{Antoniuk23}. Regarding the analytical levels, the example is
illuminating because it involved three or four levels in Fig.~\ref{fig:
  levels}.

To begin with, the incident was not only organizational but also sectoral due to
the involvement of multiple companies in the same sector. Given the potential
concentration of the energy sector companies to some particular regions in
Denmark, the incident could be perhaps interpreted to involve also the regional
level. In any case, the sectoral level manifested itself also in the
investigation conducted by \citet{SektorCERT23}, a non-profit CSIRT composed by
Danish critical infrastructure companies and operators. The investigation
revealed that the compromises were conducted by exploiting a vulnerability in a
firewall product.\footnote{~CVE-2023-28771.} Furthermore, an update had been
available but the energy sector companies had not patched their firewalls for a
reason or another. During the incident management, the issue escalated further
to the national level.

Among other things, \citet{CFCS23}, the authoritative CSIRT in Denmark, raised
the threat level against the Danish energy sector to the very high level
category. However: while multiple levels were involved, the incident is not yet
something that could be interpreted as a cyber security crisis according to the
EU's laws. This point will become clear in Subsection~\ref{subsec: events} that
elaborates different incident types present in the recently enacted EU laws.

\subsection{The EU's Strategy}\label{subsec: strategy}

\begin{figure}[th!]
\centering
\includegraphics[width=\linewidth, height=4cm]{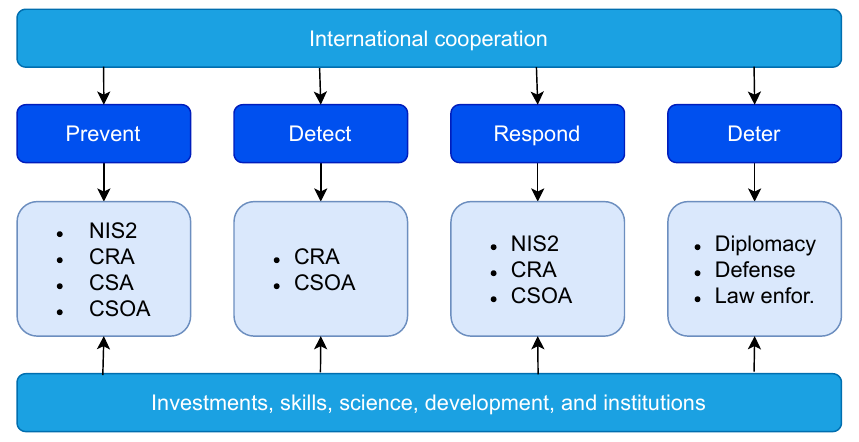}
\caption{The EU's Current Cyber Security Strategy in a Nutshell (adopted from \citealt{EC25a})}
\label{fig: strategy}
\end{figure}

The EU's new cyber security strategy can be summarized in the form of
Fig.~\ref{fig: strategy}. Three of the strategy's \text{pillars---prevention},
detection, and \text{response---resemble} rather similar phases in incident
management frameworks. For instance, the NIST's noted framework is structured
around preparation, detection and analysis, containment, eradication, and
recovery, and post-incident phases~(\citealt{NIST12}; cf.~also
\citealt{Mitropoulos06}). Rather analogously, \citet{ENISA24a} builds upon
prevention, preparedness, and response and recovery phases. Another example
would be prepare, detect and recover, and learn~\citep{Jaatun09}. A further
point is that the EU's strategy is structured around specific laws, which are
also supported by funding instruments, education, research, development, and
innovation projects, and specific governance bodies, some of which are new.

Before continuing to enumerate the laws covered, it should be emphasized that
there are many further EU laws explicitly or implicitly related to cyber
security, some of which are sector-specific and some others of which overlap
with each other~\citep{EC25b, FischerHubner21, Ruohonen25RE}. Against this
backdrop, it is no wonder that regulatory fragmentation and complexity have been
a source of commonly expressed criticism~\citep{Alexopoulos25,
  Ruohonen24I3E}. In any case, it should be emphasized that the specific laws
considered are particularly relevant for the paper's topic. In other words, as
will elaborated in Section~\ref{sec: incident types}, the laws also help to
understand how a transcendence from an incident to a unionwide crisis is legally
perceived and structured.

Of the four laws abbreviated in Fig.~\ref{fig: strategy}, particularly important
are the second network and information security (NIS2) directive~\citep{EU22a}
and the Cyber Resilience Act (CRA) regulation~\citep{EU24a}. Although a concept
of essential and important entities (EAIEs) is used in the NIS2 directive, the
directive is essentially about critical infrastructure protection. Overall, the
directive builds upon a sectoral approach; numerous sectors are enumerated as
carrying ``critical societal or economic activities'', to quote from the NIS2's
Article~2(2)(b). This quotation aligns with the notion of societal cyber
security noted in the introduction. The actual sectors specified range from
traditional critical infrastructure sectors, such as energy and transport, to
banking, healthcare, finance, drinking and waste water management, the
Internet's core infrastructure, and space technologies.

In contrast, the CRA regulation is a product-specific law. While in the NIS2
directive criticality as a concept is tied to sectors, as has been typical in
the critical infrastructure domain, in the CRA products are categorized
according to their real or perceived criticality to cyber
security~\citep{Ruohonen25JSS}. Regardless of these categorizations and with a
small number of explicitly excluded products, practically all products must
comply with the CRA's essential cyber security requirements. As these
requirements have already been considered in detail~\citep{Ruohonen25RE},
including with respect to new obligations regarding vulnerability coordination
and disclosure~\citep{Ruohonen24IFIPSEC}, in what follows, the CRA is only
discussed with respect to its mandates for incident reporting. Having said that,
it is worth remarking that the CRA's essential requirement about preferably
automated security updates would have prevented the Danish energy sector case
noted earlier in Subsection~\ref{subsec: danish}, and, furthermore, some of the
essential requirements contain also elements aligning with the technical
detection and prevention of incidents.

Regarding the CRA more generally, also the older Cybersecurity Act (CSA) can be
mentioned~\citep{EU19a}. While it strengthened the role and mandates of ENISA,
it also introduce a common cyber security certification scheme for information
technology products. This certification scheme aligns with the CRA in that
conformance can be attained also (but not only) through certification. Given
that standards are also an important part of the CRA and compliance with it, the
regulation seems to have at least partially answered to criticism from the
industry and practitioners~\citep{FischerHubner21}. Finally, recently a new
Cyber Solidarity Act (CSOA) regulation was agreed upon~\citep{EU25a}. As can be
seen from Fig.~\ref{fig: strategy}, it covers aspects from all the three pillars
noted earlier.

A couple of additional remarks are in order before continuing to elaborate the
laws and their implications in detail. The first remark is that deterring of
cyber security threats appears in Fig.~\ref{fig: strategy}. While this topic is
beyond the paper's scope, it can be noted that deterrence has long been debated
in the literature~\citep[among many others]{Goodman10}, some authors having
taken a critical stance about the working of deterrence in the cyber security
context~\citep[among them][]{Soesanto21}. The second remark is that all of the
laws are more or less risk-based, some explicitly and some only implicitly. In
particular, the NIS2 directive's Article~21 obliges EAIEs to carry out
comprehensive risk analyses. These should be done by following a so-called
``all-hazards'' approach, which is also familiar from the academic
literature~\citep{Ayyub07, Izumi24}. According to NIS2, the approach means that
anything and everything from incident handling, backups, and supply-chain
security to security awareness campaigns and business continuity should be
assessed. The CRA too is a risk-based regulation; the essential cyber security
requirements should be prioritized, designed, implemented against particular,
product-specific risks identified.

\section{Incident Types}\label{sec: incident types}

\subsection{Events}\label{subsec: events}

\begin{figure}[th!]
\centering
\includegraphics[width=\linewidth, height=4.5cm]{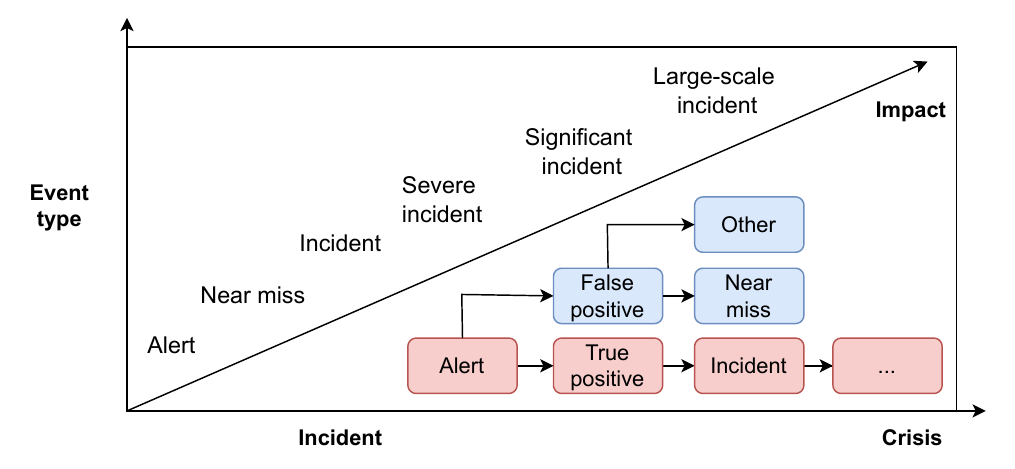}
\caption{An Event Type Hierarchy}
\label{fig: events}
\end{figure}

Events, including alerts, whether from intrusion detection systems, other
monitoring systems, or somewhere else, are important building blocks for
incident management~\citep{NIST12}. These also allow understanding how incidents
may transform into crises according to the EU's jurisprudence. Thus,
Fig.~\ref{fig: events} displays the core event types in the new EU laws. The
figure can be elaborated by moving from the bottom to the top, from alerts to
incidents, severe incidents, and beyond. In line with the previous discussion,
at least the last incident type shown is analytically already on the side of
crises.

As was remarked in Subsection~\ref{subsec: strategy}, the CRA's essential cyber
security requirements for products contain elements that may (or should) foster
the detection cyber security incidents. For instance, the requirements include a
monitoring functionality for a product's internal activity, including with
respect to authorizations and integrity guarantees~\citep{Ruohonen25RE}. In
addition, as seen also from the earlier Fig.~\ref{fig: strategy}, the CSOA
regulation is relevant too with respect to incident detection. In particular,
the regulation envisions a development of a large-scale, pan-European network of
cyber security alert systems for coordinated incident detection and improved
situational awareness capabilities. While these systems are labeled as ``cyber
hubs'' in the regulation, they are essentially what security operation centers
(SOCs) and threat intelligence systems in general are about. The network is
planned upon voluntary pooling of national alert systems into large cross-border
systems. The national systems may include not only those maintained by
authoritative CSIRTs but also those operated by private sector companies.

\subsubsection{Near Misses}\label{subsec: near misses}

An alert can be a false positive or a true positive. Although the NIST's
definition noted in the introduction covers both, for the present purposes, it
makes sense to frame incidents only toward true positives. Regarding false
positives, the EU's recent laws introduce a concept of near misses. A near miss
is defined in the NIS2's Article 6(5) to mean ``an event that could have
compromised the availability, authenticity, integrity or confidentiality of
stored, transmitted or processed data or of the services offered by, or
accessible via, network and information systems, but that was successfully
prevented''. Thus, within the context of incident management, the concept can be
seen to be about false positives; about alerts that were successfully prevented
from becoming actual incidents (true positives). In any case, reporting about
near misses is voluntary. According to Article~15(2) in the CRA regulation,
manufacturers and other parties, whether natural or legal persons, may
voluntarily report near misses either to a national authoritative CSIRT or
ENISA. Likewise, according to Article~30 in the NIS2 directive, EAIEs may
voluntary report near misses to a authoritative CSIRT or some related public
sector authority.

Voluntary reporting seems sensible already due to overhead and related
reasons. Because alerts are voluminous due to various monitoring systems
deployed by organizations, most of them are false positives, and
interoperability is a typical issue with the monitoring systems and their
reporting formats~\citep{Alahmadi22, Zibak22}. Analogously to abuse reporting in
the Internet~\citep{Jhaveri17}, a signal to noise ratio is often low. Against
these backdrops, a mandatory reporting of near misses would have presumably
required large financial investments and considerable coordination between
various stakeholders involved, including the authoritative European CSIRTs and
producers of cyber security monitoring solutions. The issues with false
positives in the cyber security context are well-recognized in the literature
also more generally; not only is a minimization of them generally difficult but
also a cost of handling them is often significant~\citep{Bhatt14}. However, the
EU laws do not say anything how organizations should actually demarcate between
false and true positives, and whether there are any consequences from
accidentally reporting a false positive as a true positive or the other way
around.

\subsubsection{Incidents}\label{subsec: incidents}

Also reporting of most (but not all) true \text{positives---that} is, according
to the terminology adopted, \text{incidents---is} voluntary (see Table~\ref{tab:
  laws}). According to the NIS2's Article 6(6), an incident is defined as ``an
event compromising the availability, authenticity, integrity or confidentiality
of stored, transmitted or processed data or of the services offered by, or
accessible via, network and information systems''. When compared to the earlier
definition for near misses, an incident is thus something that has actually
compromised data or services. While the definition aligns with classical
framings~\citep{Brownlee98}, there is a difference to the NIST's definition
noted in the introduction. Although availability, authenticity, integrity, and
confidentiality may well belong to an organization's cyber security policy, the
NIS2 only notes, in Article~24, cyber security policies with respect to open
source software associations. In any case, according to the NIS2's Article~30,
reporting of these ``conventional'' incidents is voluntary.

\begin{table*}[th!b]
\centering
\begin{threeparttable}
\caption{References for Legal Definitions and Reporting Obligations}
\label{tab: laws}
\begin{tabular}{llll}
\toprule
Type & Discussion & Definition & Reporting \\
\hline
Near miss$^a$ & Subsection~\ref{subsec: near misses} & Article 6(5) in the NIS2 directive & Voluntary \\
Incident$^b$ & Subsection~\ref{subsec: incidents} & Article 6(6) in the NIS2 directive & Voluntary \\
Severe incident & Subsection~\ref{subsec: severe incidents} & Article 14(3) in the CRA & Mandatory \\
Significant incident & Subsection~\ref{subsec: significant incidents} & Article 23(3) in the NIS2 directive & Mandatory \\
Large-scale incident & Subsection~\ref{subsec: large-scale incidents} & Article 6(7) in the NIS2 directive & Mandatory \\
\hline
Actively exploited vulnerabilities & Subsection~\ref{subsec: severe incidents} & Article 3(42) in the CRA & Mandatory$^c$ \\
\bottomrule
\end{tabular}
\begin{tablenotes}
\begin{scriptsize}
\vspace{3pt}
\item{$^a$~The same article in the NIS2 directive is explicitly referenced in the
  CRA's Article~3(45).}
\vspace{3pt}
\item{$^b$~The same article in the NIS2 directive is explicitly referenced in
  the CRA's Article~3(43). Although the so-called critical entities resilience
  (CER) directive \citep{EU22b} was omitted from the paper's scope
  (cf.~Fig.~\ref{fig: strategy}), an incident is defined also in its
  Article~2(3). For a reason or another, however, the CER's definition is not
  identical with the NIS2 directive's definition.}
\item{$^c$~Though, there is an option to delay
  reporting~\citep{Ruohonen24IFIPSEC}.}
\end{scriptsize}
\end{tablenotes}
\end{threeparttable}
\end{table*}

\subsubsection{Severe Incidents}\label{subsec: severe incidents}

When moving a step upward in the event hierarchy (see Fig.~\ref{fig: events}),
the CRA regulation (but not the NIS2 directive) uses a concept of severe
incidents. It is mandatory to report such incidents to both a national
authoritative CSIRT and ENISA according to the CRA's Article~14(3). A~twofold
definition for severe incidents is given in the regulation's Article~14(5). The
first part emphasizes negative impacts \textit{and} potential negative impacts
upon a product's ability to protect the availability, authenticity, integrity,
or confidentiality of data or services, whereas the second part emphasizes an
execution \textit{and} a potential execution of malicious code. It remains to be
seen how manufacturers will interpret the demarcation between the impacts and
executions on one hand and the potential impacts and executions on the other
hand.

Another important point is that the CRA is about manufacturers of information
technology products, not about operators of such products, including EAIEs
covered by the NIS2 directive. With some relaxations, manufacturers are also
mandated to report about actively exploited
vulnerabilities~\citep{Ruohonen24IFIPSEC}. These are archived to a new European
vulnerability database maintained by \citet{ENISA25b}. In any case, the
divergence between the scopes of the two EU laws reiterates a criticism
expressed both in research and industry about a lack of an explicit legal
synchronization between different reporting obligations imposed by different
laws~(\citealt[p.~49]{BusinessEurope25}; \citealt{FischerHubner21};
\citealt{Ruohonen24ISJGP}). In particular, in case a manufacturer reports to a
national authoritative CSIRT and ENISA about a severe incident, either the given
CSIRT or ENISA, or both, should presumably consequently report to all EAIEs and
others using the manufacturer's given product to which the severe incident
applies. Again, it remains to be seen how such consecutive reporting will work
in practice.

\subsubsection{Significant Incidents}\label{subsec: significant incidents}

The remaining two incident types are specified in the NIS2 directive. According
to the directive's Article~23(3), a significant incident either ``has caused or
is capable of causing severe operational disruption of the services or financial
loss for the entity concerned'', or it ``has affected or is capable of affecting
other natural or legal persons by causing considerable material or non-material
damage''. The again twofold definition warrants three brief remarks.

The first remark is related to the earlier point about potential effects; the
wording ``capable of'' yet again entails a demarcation problem between actual
significant incidents and those merely conveying a probability of such
incidents. Presently, no guidance is available upon this
demarcation~\citep{Busetti25}. The second remark is related to the wording about
``severe operational disruption''. Here, it would seem that anything from severe
denial of service attacks to severe ransomware attacks are in the scope. The
definition's emphasis of severe financial losses further underlines ransomware
attacks, although also other so-called destructive cyber attacks are likely in
the NIS2's scope. The third remark is about the emphasis about ``considerable
material or non-material damage'' to natural persons. Therefore, it has been
suspected that also severe personal data breaches are in the directive's
scope~\citep{Ruohonen25RE}, but a definite conclusion can likely be given only
after guidelines have been released or enforcement has occurred.

While keeping these and other potential interpretation issues in mind, including
regarding the rich role of adjectives and other qualifying words in law
\citep{Bertoldi07}, EAIEs must report about significant incidents to a national
authoritative CSIRTs or, where applicable, other public authorities, as well as
other entities concerned according to the NIS2's Article~23(1).\footnote{~All
laws considered contain numerous qualifying words~\citep{Ruohonen25ISTb}. For
the present purposes, particularly the many adjectives present in the
definitions listed in Table~\ref{tab: laws}, such as severe, significant, and
considerable, convey also semantic meanings. However, a contextual
interpretation is only possible once guidance is available or enforcement
actions have occurred.} Thus, broadly speaking, unlike ``conventional''
incidents, significant incidents fall into a category of mandatory reporting by
EAIEs. According to guidance from \citet{ENISA25c}, EAIEs should also carefully
evaluate and document all significant incidents and adjust their risk analysis
results accordingly.

\subsubsection{Large-Scale Incidents}\label{subsec: large-scale incidents}

Finally, there is the notion about large-scale cyber security
incidents. A~large-scale incident is defined in the NIS2 directive's
Article~6(7) as an incident which either (a)~exceeds a handling capacity of a
single member state or which (b)~has a significant impact on at least two member
states. Regarding incident management frameworks, the first part of the
definition resembles a concept of escalation at the organizational level; an
organizational CSIRT or some related security team cannot get an incident under
control on its own~\citep[cf.][]{Jaatun09}. Another point is that the
\citeauthor{EC25b}'s \citeyearpar[p.~12]{EC25b} noted recent proposal clarifies
that these large-scale incidents are what is meant by a cyber security crisis in
the EU context. While keeping the earlier remarks about terminology in mind, a
cyber security crisis is thus something that moves from the national level to
the supranational level in Fig.~\ref{fig: levels}. Also the NIS2's recital 69
clarifies this point by noting that large-scale incidents ``may escalate and
turn into fully-fledged crises''. Despite such an escalation potential and
interestingly enough, NIS2 does not explicitly say about reporting obligations
of large-scale incidents, but because these must be something that also satisfy
the definition for significant incidents, reporting can be interpreted as
mandatory.

The CSOA's cyber security alert systems noted in Subsection~\ref{subsec: events}
are also related to large-scale cyber security incidents. According to the
regulation's Article~7, particularly the envisioned cross-border alert systems
are seen as relevant for detecting large-scale incidents and sharing information
about them. There are also a couple of other important points to make about the
relation between large-scale incidents and the CSOA regulation.

The first point is about the exceeding of a member state's handling capacity
used in the definition for a large-scale cyber security incident. Here, the
CSOA's Article~14 specifies an establishment of a specific cyber security
``reserve'' for handling large-scale incidents. The reserve is composed by
national public sector representatives as well as other trusted
parties. According to Article~15, a member state in need may then request help
from the reserve to recover from both significant and large-scale cyber security
incidents. Furthermore, the CSOA's Article~19 specifies that help can be
requested also by countries outside of the EU, provided that they are taking
part in the EU's Digital Europe Programme (DEP), a funding instrument for
digital technologies and digitalization in general.

The second point is that the CSOA specifies a new cyber security incident review
mechanism. Its scope is restricted to significant and large-scale
incidents. ENISA and a network of authoritative CSIRT are specified as the
responsible parties for the actual reviews according to the regulation's
Article~21(1). In line with the post-incident activities in the NIST's incident
management framework~\citep{NIST12}, the reviews are used for improving the
union's and the member states' cyber security practices. Despite an incident's
or a crisis' negative consequences, there is always also something to learn from
them.

\subsection{Deadlines}

\begin{figure}[th!]
\centering
\includegraphics[width=\linewidth, height=3cm]{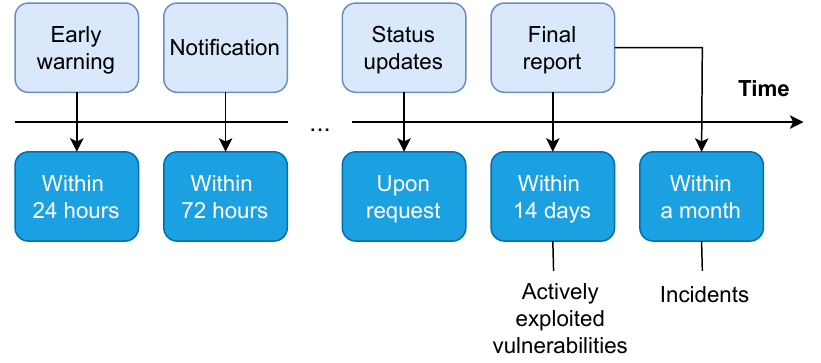}
\caption{Reporting Deadlines for Actively Exploited Vulnerabilities and Severe,
  Significant, and, by implication, Large-Scale Incidents}
\label{fig: deadlines}
\end{figure}

There are strict deadlines for incidents that are mandatory to report. When
comparing the CRA's Article~14 and the NIS2 directive's Article~23, the wordings
about the deadlines are not verbatim similar but their meaning is more or less
the same.\footnote{~Although data protection was framed out from the paper's
scope~(see Subsection~\ref{subsec: framings}), it is still worth mentioning that
in case personal data is processed, there is an overlap with the General Data
Protection Regulation~(GDPR) \citep{EU16}, as also recognized in existing
research~\citep{Ruohonen25RE}. That is, according to the GDPR's Article~33, most
personal data breaches must be communicated to data protection authorities
within 72 hours. In addition, some sector-specific laws, such as the one for the
resilience of financial services~\citep{EU22c}, impose further incident
reporting obligations.}  For this reason, the deadlines can be summarized in the
form of Fig.~\ref{fig: deadlines}. As was noted in the preceding
Subsection~\ref{subsec: large-scale incidents}, the deadlines for reporting
about large-scale incidents are implicit but follow from their inevitable
connection to significant incidents. Another clarifying remark needed is that
the CRA's notion about actively exploited vulnerabilities is not about incidents
\textit{per~se} but can be discussed alongside them because the reporting
deadlines are similar.

Thus, severe, significant, and large-scale incidents should be initially
reported to an authoritative CSIRT, or some other competent authority, no later
than twenty-four hours after becoming aware of them. Then, a more detailed
incident report should be delivered no later than seventy-two hours after
becoming aware of a given incident. The report delivered should include an
initial assessment about a given incident, including its severity and impact, as
well as, where applicable, any indicators of a compromise. After these two
reports, an authoritative CSIRT may request status updates at any time. A final
report should be delivered within a month. In addition to severity and impact,
the final report should contain a root cause analysis, any applied and ongoing
mitigation measures, and, where applicable, an assessment about any cross-border
impacts. The cross-border impact assessment again signifies the relation to
large-scale incidents. Regarding actively exploited vulnerabilities, the CRA's
Article~14(2)(c) specifies that a final report should be delivered already no
later than fourteen days. It should describe a given vulnerability's severity
and impact, information about potential exploitation, and details about security
updates or other corrective measures made available.

Finally, it should be emphasized that while these reporting obligations and
their deadlines are only toward public authorities, also other parties typically
need to be informed, including further external parties, such as stakeholders
and business partners, and internal parties, such as an organization's
management and, where applicable, its employees~\citep{Kulikova12}. Thus,
existing incident management plans at the organizational level presumably
require or benefit from updating; a synchronization with the law-imposed
obligations might be a good idea. Due to the various reporting obligations and
strict deadlines, an analogous point applies on the public sector side;
so-called incident triaging~\citep{ENISA10} likely requires planning.

\section{Crisis Management}\label{sec: crisis management}

\subsection{Governance Units}

The management of cyber security crises is generally challenging in the EU due
to the involvement of at least two member states, given the definition for a
large-scale cyber security incident noted in Subsection~\ref{subsec: large-scale
  incidents}. Given the analytical levels discussed in Subsection~\ref{subsec:
  levels}, the management is about the national level and the supranational
level. In terms of the former, the term transnational could be used to
characterize the horizontal interstate management and coordination between the
member states involved. While governments may be involved too in a case of
particularly severe crises, the term intergovernmental might be too strong or
even misleading because the actual day-to-day crisis management would not likely
involve governmental representatives in most cases. The NIS2 directive also
established new crisis management bodies that are not explicitly and directly
about the member states' central governments. These also coordinate vertically
toward units operating at the supranational level.

In general, the NIS2 directive specifies cyber security crisis management to
occur through three governance bodies:
\begin{enumerate}
\item{The first is the authoritative network of European CSIRTs upon with the
  EU's whole cyber security framework was initially
  built~\citep{Ruohonen16GIQ}. As the CSIRT network extends beyond Europe, the
  international level may be involved too, as also indicated by Fig.~\ref{fig:
    strategy}. In general, NIS2 strengthens the European CSIRT network further,
  obliging also the authoritative CSIRTs to carry out further tasks.}
\item{The second is a specific NIS2 cooperation group. However, as specified in
  the NIS2's Article~14(4), its tasks are largely on the political side instead
  of the operational crisis management side. Among other things, the group is
  specified to help at formulating further policies, exchanging best practices
  and viewpoints, including regarding sectoral implementations, carrying out
  risk analyses, providing strategic guidance, and meeting with private sector
  stakeholders.}
\item{The third is a new European cyber crisis liaison organisation network
  (EU-CyCLONe). As specified in the NIS2's Article~16, unlike the cooperation
  group, EU-CyCLONe is on the operational side; in particular, it is explicitly
  tasked to manage large-scale cyber security incidents. In general, the
  management involves similar phases that were noted in Subsection~\ref{subsec:
    strategy}, among them preparedness and response. The EU-CyCLONe network is
  specified to also coordinate with the network of authoritative European
  CSIRTs.}
\end{enumerate}

The political side of crisis management deserves a further comment. This side is
seen in the composition of the NIS2 cooperation group. As specified in the
NIS2's Article~14(3), it is composed of representatives of the member states,
the EC, and ENISA, but also other public sector authorities may be present,
including the European External Action Service (EEAS) who is responsible for the
EU's foreign policy. While the activities of the EEAS were framed to outside of
the paper's scope via the discussion in Subsection~\ref{subsec: framings}, it is
still worth emphasizing that cyber security crisis management in the EU may
involve also diplomacy and related foreign policy activities, as also indicated
by the summary in Fig.~\ref{fig: strategy}. Analogously, the NIS2 directive's
Article~16(3)(d) specifies that the EU-CyCLONe network should also ``support
decision-making at political level''.

Given these elaborations, it is understandable that some criticism has been
levied about increased administrative complexity, fragmentation, and
bureaucratization of the EU's cyber security governance model in
general~\citep{Ruohonen24ISJGP, Ruohonen24I3E}. Such criticism aligns with a
broader branch of research on the ``bureau-politics'' of crisis management
\citep{Rosenthal91}, including potential ``turf wars'' between governance
bodies~\citep{Finke20, Senninger21}. The COVID-19 pandemic provides an excellent
reference point in this regard~\citep[cf.][]{Lipscy20, Zahariadis23}.  This
research branch is relevant to note because rapid responses and timeliness in
general are important particularly in the cyber security context. As already the
deadlines in Fig.~\ref{fig: deadlines} demonstrate, rapid responses are required
from EAIEs and others, but it remains generally unclear how fast and well the
responses travel through the bureaucracy and politics at the receiving end.

\subsection{Management at the Supranational Level}

\begin{figure*}[t!]
\centering
\includegraphics[width=13cm, height=5cm]{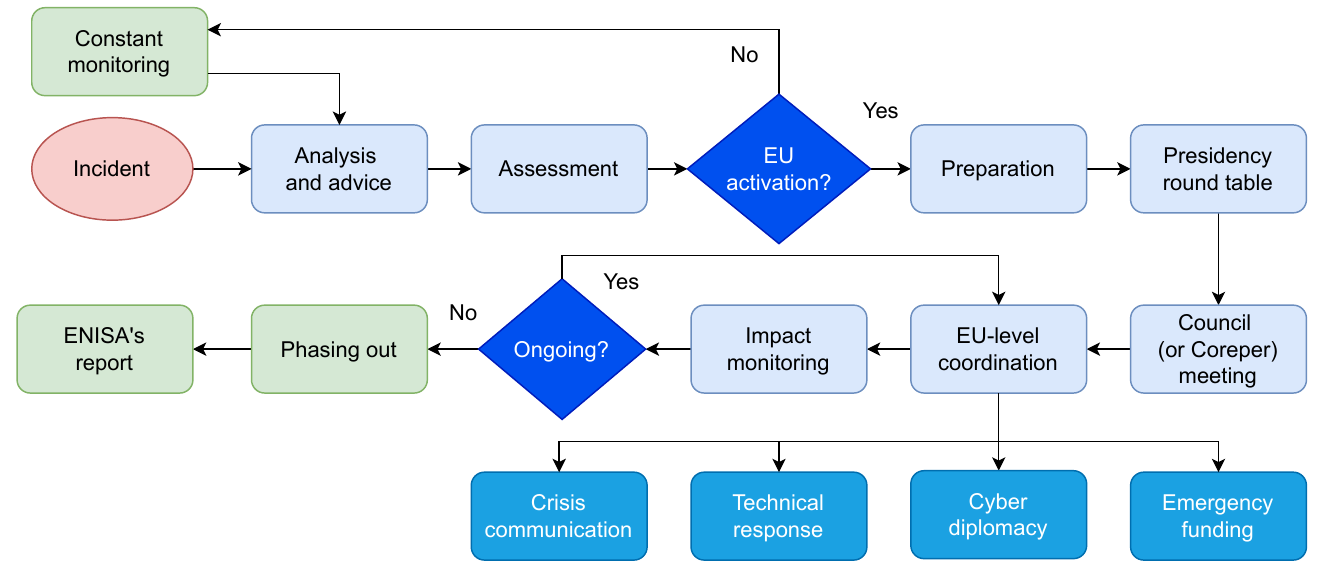}
\caption{Cyber Security Crisis Management at the EU-Level (adopted and modified
  from \citealt[Fig.~2]{EU17})}
\label{fig: eu-level}
\end{figure*}

The CSIRT and EU-CyCLONe networks as well as the NIS2 cooperation group operate
both at the national level and the supranational level. At the former level,
these allow horizontal coordination between the member states; at the latter
level, these enable the member states to coordinate with EU-level institutions,
among them the EC and ENISA. As was noted in Subsection~\ref{subsec:
  supranational level}, the Council is on the side of intergovernmental
governance, and within the Council, the member states have permanent
representatives through the Committee of the Permanent Representatives of the
Governments of the Member States to the European Union---or Coreper in
short. With these clarifications, a high-level crisis management framework at
the EU-level can be illustrated in the form of Fig.~\ref{fig: eu-level}.

The process starts when an early warning or a later incident notification is
received, possibly from an EAIE or a larger set of EAIEs, through an
authoritative CSIRT or multiple authoritative CSIRTs. For the present purposes,
the phase after the initial assessment, the EU activation phase, is the crucial
one; during this phase, the two part definition for a large-scale cyber security
incident should be likely assessed. Politically, the activation is done by the
rotating Presidency of the Council after consulting the EC and the EEAS' high
representative~\citep{EU18}. As noted in the CSOA's Article~20, the political
activation procedure is similar to requesting help from the new cyber reserve as
well as those used for natural and other disasters requiring civil
protection. Technically, however, the available documentation is unclear about
the actual incident analysis. The EEAS, EC, ENISA, CSIRT and EU-CyCLONe networks
are mentioned alongside Europol, the EU's own CSIRT (known as CERT-EU), and even
the EU's intelligence and satellite institutions~\citep{EC25b, EU17}. Given that
the political activation depends on, or should depend on, the adequacy and
correctness of a technical incident analysis, already the amount of potential
actors is enough to reiterate the earlier point about administrative complexity
and its possible consequences. Nor is it possible to deduce what might be a cost
of a false positive.

Assuming that political activation is done, a further bottleneck might be the
subsequent phases. As seen from Fig.~\ref{fig: eu-level}, there are preparations
and meetings before the actual EU-level coordination starts looping. If a
response time is short, these phases may cause undesirable delays. In addition
to a technical response, including the containment, eradication, and recovery
tasks in the NIST's framework~\citep{NIST12}, the EU's crisis management plans
include also considerations about funding needs, diplomatic responses, and
crisis communication. In addition to academic research~\citep{Kulikova12}, the
role and importance of crisis communication have recently been emphasized also by
\citet{ENISA24a} who further stressed a need to develop clear technical
indicators and decision mechanisms for activating the EU machinery.

\subsection{A Hypothetical Scenario}\label{subsec: scenario}

\subsubsection{Setup}

\begin{figure}[th!]
\centering
\includegraphics[width=\linewidth       , height=4.5cm]{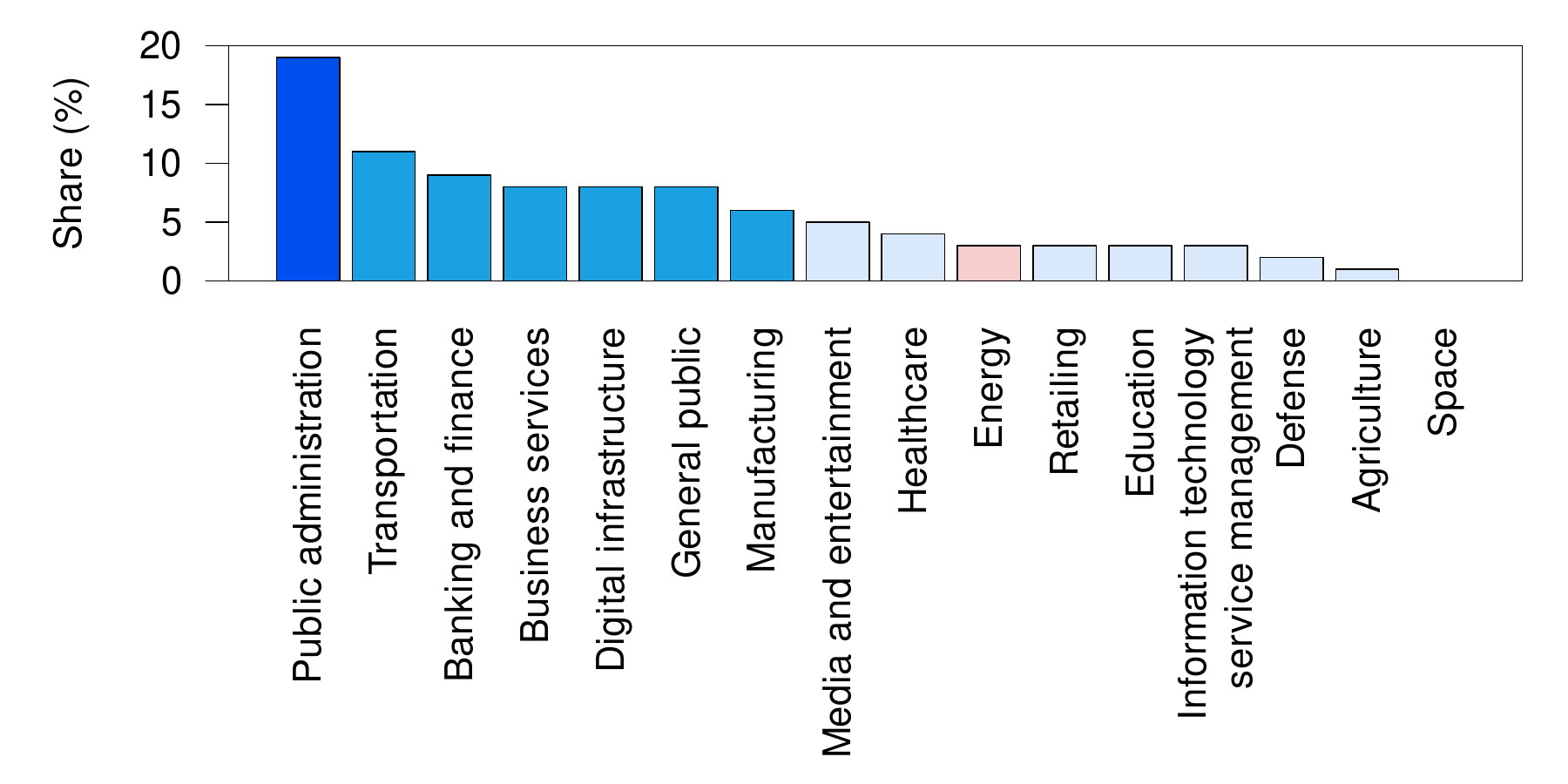}
\caption{Shares of Reported Incidents in Selected Sectors Between July 2023 and June 2024 in the European Union (adopted from \citealt[Fig.~5 on p.~18]{ENISA24b})}
\label{fig: incidents}
 \end{figure}

A brief hypothetical scenario can be presented by extending the real-world case
briefly described in Subsection~\ref{subsec: danish} toward a cyber security
crisis. Thus, the context is again the energy sector. Although the European
energy sector has not seen frequent incidents (see Fig.~\ref{fig: incidents}),
supposedly because of the sector's cyber security maturity~\citep{ENISA24b}, the
energy sector is categorized as critical in the NIS2 directive. The energy
sector also continuously ranks high in terms of criticality
assessments~\citep{ENISA25a}. As will be seen, the sector's criticality
partially stems from potential cascading effects. To this end and other ends,
the energy sector has also been a topic of exercises conducted by the national
and EU-level authorities~\citep{ENISA24b}. To further frame the hypothetical
scenario toward the real-world case, the context can be restricted to the Nordic
countries. The electricity grids in these countries are mostly connected and
synchronized.\footnote{~Though, Iceland and parts of Denmark are disconnected.}
The large synchronized Nordic smart grid further has non-synchronized links to
Germany, Poland, Lithuania, the Netherlands, the United Kingdom, and
Estonia~\citep{SK25}. When keeping in mind the definition for a large-scale
incident or a crisis (see Subsection~\ref{subsec: large-scale incidents}), the
Nordic smart electricity grid thus provides an excellent case for the
hypothetical scenario.

\subsubsection{Events}

With the previous notes, consider that a large international producer reported
both to ENISA and a national authoritative CSIRT in one Nordic country about an
actively exploited vulnerability in an industrial control system product. The
producer had reliable evidence about active exploitation in Southeast Asia but
was not aware about any actual incidents in Europe, including severe incidents
that would warrant additional mandatory reporting. The notification about the
actively exploited vulnerability was done within the 72 hour deadline. The
authoritative CSIRT started to investigate the vulnerability notification.

About a week after the initial notification, an authoritative CSIRT in Southern
Europe received a few almost identical reports about near misses caught by
intrusion detection systems deployed by two companies operating in the water
management sector, which too is defined as critical in the NIS2 directive. After
consulting the companies, the CSIRT concluded these to be noise that is typical
to intrusion detection systems. It did not notify ENISA or other coordination
bodies. In the meanwhile, the Nordic CSIRT was still investigating the actively
exploited vulnerability in collaboration with the international producer.

About two weeks after the initial notification about active exploitation outside
of Europe, another authoritative CSIRT in another Nordic country received an
urgent notification about a significant incident in the energy sector; two
relatively large regions were out of power due to a suspected cyber attack. Soon
after, a similar notification was received by the Lithuanian authoritative
CSIRT.  The two CSIRTs notified ENISA and started to investigate the incidents
through the EU-CyCLONe network. Thanks to redundancy mechanisms, the EAIEs
involved managed to solve the power cuts within about eight hours. As the
incidents were under control, the EU-CyCLONe network urged the EU-level
administration to not active the EU's cyber security crisis management framework
even though already the preliminary investigation revealed that the definition
for a large-scale incident was satisfied. Nevertheless, the coordination network
decided that a joint internal investigation was needed.

About a month after the initial notification, the internal investigation was
ready. It indicated that the power availability issues had resulted from a cyber
attack that involved tampering with control messages delivered to industrial
control systems used by the EAIEs involved. However, there had been a
significant \textit{cyber security incident} only in one Nordic country; the
Lithuanian case had occurred merely because of the disruption to the shared
grid. Furthermore, it was concluded that all previous notifications were
related. The root cause was the actively exploited vulnerability initially
reported to another Nordic authoritative CSIRT. Also the near misses in an
unrelated sector had been about the same vulnerability.

\subsubsection{The Points Raised}

\enlargethispage{1cm} 

Although only hypothetically, the scenario illustrates seven points: (1) the
potential involvement of entities outside of Europe; (2) the criticality
concept's ties not only to sectors but also to products; (3) the potential
cascading effects and relationships between confidentiality, integrity, and
availability; (4) the potential difficulties to triage, analyze, and correlate
different notification types; (5) the potential dangers involved in making
decisions about true and false positives; (6) the potential delays and related
problems in vulnerability and incident coordination; and (7) the potential
political and other uncertainties about the EU-level activation. Regarding
further scenarios, whether hypothetical or empirical, particularly the
escalation and containment dynamics require better understanding. Despite the
seventh point raised, the scenario is also a good way to conclude by answering
to RQ.1, RQ.2, and RQ.3.

\section{Conclusion}\label{sec: conclusion}

The paper's conclusion can be summarized by answering to the three RQs specified
in the introduction. Thus:
\begin{enumerate}
\item{The EU's new cyber security laws distinguish four incident types:
  ``conventional'' incidents, severe incidents, significant incidents, and
  large-scale incidents. Albeit only implicitly, the last incident type is
  equated to the concept of a cyber security crisis. The definition for such a
  crisis is twofold: \textit{a cyber security incident becomes a cyber security
    crisis either in case it exceeds a handling capacity of a single member
    state of the EU or in case it significantly affects at least two member
    states simultaneously}.}
\item{The EU's new cyber security laws significantly extend reporting
  obligations toward public sector authorities. Both operators of critical
  infrastructures and producers of many information technology products are
  mandated to report incidents; only reporting of the ``conventional'' incidents
  is voluntary. The deadlines for reporting are also strict. Furthermore, there
  are overlaps presents between the reporting obligations.}
\item{Cyber security crisis management and its governance in the EU are still
  strongly built upon national authoritative CSIRTs and their EU-level
  coordination bodies, including ENISA in particular. However, the new laws have
  also brought additional governance bodies, among them EU-CyCLONe that was
  specifically established for cyber security crisis management. In addition, a
  new staff reserve has been established for helping countries facing cyber
  security crises. Despite---or due to---these new crisis management bodies and
  recent policy proposals, it remains generally unclear how---and how
  well---crisis management works at the EU-level administration.}
\end{enumerate}

\section{Limitations}\label{sec: limitations}

The most notable limitation is related to interpretation, including a degree of
subjectivity that inevitably follows from descriptive policy research dealing
with laws and theory-building. Among other things, as was noted in
Subsection~\ref{subsec: significant incidents}, many of the qualifying
adjectives, such as significant and severe, are open to further
interpretation. However, definite interpretations are likely possible only after
guidance from the public authorities involved, including in terms of
administrative fines or other enforcement actions. As both the fines and other
enforcement actions, such as a potential withdrawal of non-compliant products
from the EU's internal market, can be severe, it is likely that some future
cases will end up also to courts whose decisions entail the final
interpretations.

Another notable limitation is related to the national adaptations and
implementations of the EU laws considered. Existing results indicate a
substantial variation already between the national transpositions of some
related EU cyber security laws~\citep{Becker25}.  In general---and despite the
new EU laws and frameworks for these~\citep{ENISA25a}, national and other
definitions for alert levels and criticality itself remain unclear,
non-harmonized, and generally vague~\citep{Alexopoulos25, Ruohonen24ISJGP,
  Ruohonen25JSS}.\footnote{~For instance, \citet[p.~12]{ENISA25a} has assessed
criticality in terms of socio-economic impacts, spillovers to other sectors and
countries, an incident's or a crisis' longitudinal persistence, and reliance on
information and communication technologies. Although sensible as such, it
remains unclear how objective such broad assessment criteria are. Many aspects
of criticality are also missing from sectoral frameworks, including the NIS2
directive~\citep{Ruohonen25JSS}.} Thus, further research is needed to better
understand the national level in Fig.~\ref{fig: levels}. As was noted in
Subsection~\ref{subsec: national regional and sectoral levels}, a divergence
between the member states can be expected.

\section{Further Work}\label{sec: further work}

The three concluding answers can be accompanied by four points about further
research. To motivate the first point, it can be noted that despite
uncertainties about actual crisis management particularly at the EU-level, it
seems the NIS2 directive has addressed a commonplace problem among many member
states whose national laws and frameworks lacked a definition for a cyber
security crisis~\citep{Boin18}. However, the NIS2's definition for a large-scale
cyber security incident might also be argued to be about too large crises in a
sense that an escalation to a national level might already be perceived as a
crisis in some severe cases. This point motivates the first avenue for further
research.

Given the overall escalation theme in incident management
research~\citep{Mitropoulos06}, as already said, (1) further theoretical and
comparative research is needed about the analytical levels in Fig.~\ref{fig:
  levels}. Regarding comparisons, a good starting point would be a comparative
examination of national laws and frameworks for cyber security crisis management
in the member states. National adaptations and implementations would also help
at addressing the interpretation limitation.  Also the paper's framings noted in
Subsection~\ref{subsec: framings} would deserve a theoretical visit. Despite the
research on hybrid threats, the relation between cyber security incidents and
crises, including their potential cascading effects, and other incidents and
crises remain arguably poorly understood and theorized. These points justify
further theoretical and conceptual research~too. Regarding comparative research
more generally, (2)~further research is also needed to compare the EU's cyber
security crisis management framework with other countries. China and the United
States would be good candidates for a comparison.

Regarding the EU-level cyber security crisis management, (3)~further research is
required also on the roles, duties, and other functions of various different
administrative units, whether national, supranational, or something
in-between. Composites such accountability, collaboration and coordination,
transparency, information sharing, decentralization and autonomy, and
responsiveness could be used to frame an
evaluation~\citep[cf.][]{Vu25}. However: even though it is easy to agree with an
argument that more empirical research is needed on incident management in
general~\citep{Tondel14}, a~problem with the NIS2's large-scale incidents is
that thus far at least publicly disclosed cases are missing. In other words, it
is difficult to properly evaluate a management of something that has not
supposedly yet happened in Europe.

Therefore, regarding empirical research, it might make sense to start from
smaller evaluations. In particular, (4)~the increased reporting obligations,
including the strict deadlines, would offer a good topic for empirical
evaluation research. Already the points raised in Subsection~\ref{subsec:
  scenario} would provide a theoretical motivation. A lot of research and other
work have also been done to help at evaluating the efficiency, accuracy, and
other aspects CSIRTs and their incident management practices~\citep[among many
  others]{Connell13, Dorofee07}. However, a broader evaluation is required also
with respect to EAIEs and others who are obliged to report about incidents or do
so voluntarily. Given the CRA's and NIS2's strict deadlines, as well as the
penalties from non-compliance, it may be that organizations and others will
report eagerly and possibly incautiously. Thus, the quality of early warnings,
incidents notifications, and particularly final reports~\citep{Busetti25},
including a rate of false positives, would provide a good research topic with
practical relevance. A related topic would involve examining whether reporting
guidelines and a systematic reporting format might increase reporting quality
and reduce noise. As it stands, neither the NIS2 directive nor the other laws
say anything about \textit{what} should be reported.

The reporting topic is generally important because regulatory obligations upon
incident management have recently been argued to work poorly in improving
organizations' cyber security postures~\citep{Patterson24}. This point serves
well to end the paper by reiterating that something should always be learned
from cyber security incidents and crises. Because the reports are likely
confidential, however, it may well be that the topic could only be studied by
indirect means, such as surveys or interviews of public authorities and EAIEs.

\section*{Acknowledgements}

The authors thank Muhammad Mohsin Hussain for helpful comments.

\balance
\bibliographystyle{apalike}

\end{document}